# The Defects Genome of Janus Transition Metal Dichalcogenides

Mohammed Sayyad [1, Γ], Jan Kopaczek [2], Carmem M. Gilardoni [3], Weiru Chen [4], Yihuang Xiong [4], Shize Yang [5], Kenji Watanabe [6], Takashi Taniguchi [7], Robert Kudrawiec [2], Geoffroy Hautier [4], Mete Atatüre[3] and Sefaattin Tongay [1, Γ]

[1] Materials Science and Engineering, School for Engineering of Matter, Transport and Energy, Arizona State University, Tempe, Arizona, AZ 85287, United States of America.

[2] Department of Semiconductor Materials Engineering, Faculty of Fundamental Problems of Technology, Wroclaw University of Science and Technology, Wybrzeże Stanisława Wyspiańskiego 27, 50-370 Wroclaw, Poland.

[3] Cavendish Laboratory, University of Cambridge, J.J. Thomson Avenue, Cambridge, CB3 0HE, United Kingdom.

[4] Thayer School of Engineering, Dartmouth College, Hanover, New Hampshire, NH 03755, United States of America.

[5] Aberration Corrected Electron Microscopy Core, Yale University, New Haven, Connecticut, CT 06516, United States of America.

[6] Research Center for Functional Materials, National Institute for Materials Science, Tsukuba 305-0044, Japan.

[7] International Center for Materials Nanoarchitectonics, National Institute for Materials Science, Tsukuba305-0044, Japan.

Γ *Corresponding Authors: sefaattin.tongay@asu.edu, msayyad@asu.edu*

**Abstract**

Two-dimensional (2D) Janus Transition Metal Dichalcogenides (TMDs) have attracted much interest due to their exciting quantum properties arising from their unique two-faced structure, broken-mirror symmetry, and consequent colossal polarisation field within the monolayer. While efforts have been made to achieve high-quality Janus monolayers, the existing methods rely on highly energetic processes that introduce unwanted grain-boundary and point defects with still unexplored effects on the material's structural and excitonic properties Through High-resolution scanning transmission electron microscopy (HRSTEM), density functional theory (DFT), and optical spectroscopy measurements; this work introduces the most encountered and energetically stable point defects. It establishes their impact on the material's optical properties. HRSTEM studies show that the most energetically stable point defects are single ($V_S$ and $V_{Se}$) and double chalcogen vacancy ($V_S$–$V_{Se}$), interstitial defects ($M_i$), and metal impurities ($M_W$) and establish their structural characteristics. DFT further establishes their formation energies and related localized bands within the forbidden band. Cryogenic excitonic studies on h-BN-encapsulated Janus monolayers offer a clear correlation between these structural defects and observed emission features, which closely align with the results of the theory. The overall results introduce the defect genome of Janus TMDs as an essential guideline for assessing their structural quality and device properties.

## Introduction

Named after the Roman god, Janus transition metal dichalcogenides (TMDs) are a novel class of atomically thin two-dimensional (2D) materials[1]. Unlike classical TMDs[2] with the chemical formula $MX_2$ (M=Mo, W and X=chalcogens), Janus TMDs are characterized by the 'broken' mirror symmetry, resulting from top and bottom surfaces containing different chalcogen atoms. As such, they can be referred to in $M_X^Y$ notation to indicate the bottom chalcogen atom type with the subscript X and the top chalcogen atom type as Y superscript[3,4]. Theoretical and experimental studies to date have shown that the broken mirror symmetry along with permanent dipole lends Janus TMDs with exotic effects such as Rashbatronics[5-7], colossal piezoelectricity[8,9], long-lived dipolar excitons[10], enhanced Dzyaloshinskii-Moriya Interaction[11], skyrmion formation[12,13], and polarisation driven photocatalytic water splitting[14].

So far, much effort has focused on synthesizing 2D Janus materials with high structural, electronic, and excitonic quality, creating strain-engineered high-order structures such as Janus nanoscrolls reported recently [15,16], as well as finding effective ways to expand the library of 2D Janus materials[17-20]. These studies readily opened routes for exploring complex many-body physics in these materials systems[10,21-27]. Despite efforts to realize structurally perfect Janus layers, the existing methods still involve either plasma or pulsed laser processing to transform classical TMDs into 2D Janus layers[3,4,18-20,28]. These non-equilibrium and highly energetic processes introduce unintentional point and grain boundary (GB) defects. However, the relationship between the material's structural characteristics and intrinsic electronic and optical responses remains underexplored[29-31]. For example, understanding the behaviour of defects in 2D materials[32-41] has enabled the realization of many quantum effects, including valley coherence[42-46], spin polarised Landau levels[47,48], valley Zeeman effect[49], ferromagnetic ordering[50,51], the realization of single quantum emitters[52-54]. Similarly, the fundamental understanding of the effects of defects on the structural and material properties of Janus TMDs holds the key to their high-quality manufacturing, understanding their quantum properties, and assessing device properties of 2D Janus TMDs.[15,55-58]

Here, our work introduces the first comprehensive investigation of impurities and defects in Janus monolayers[28] and correlates them to their excitonic properties. Employing ultra-low electron dose HR-STEM on Janus $W_{Se}^{S}$ and $Mo_{Se}^{S}$ monolayers, we identify and structurally characterize the most common point defects and grain boundaries. Using Density Functional Theory (DFT), the results establish the defect formation energy and identify the most thermodynamically favourable defect states. Cryogenic photoluminescence spectroscopy of hBN encapsulated Janus TMDs enabled us to identify three isolated bound exciton emission lines and correlate these spectral signatures to point defects [36,59,60,61]. Overall results, interpreted DFT simulations, correlate imperfections to the optical properties of 2D Janus TMDs.

**Results**
2D Janus synthesis was carried out using established growth techniques in the literature, starting from large area chemical vapour deposited (CVD)[15,17,21,28] $WSe_2$ and $MoSe_2$ or exfoliated sheets from CVT grown vdW crystals[15,28,33]. These classical monolayers were transformed into Janus TMDs using plasma-based techniques [15,17,28] at high excitonic qualities. More specifically, classical TMDs were positioned in a custom-designed quartz chamber, retrofitted with a viewing window that couples it into Raman and PL spectrometers for in-situ monitoring capabilities to ensure high-quality Janus layer synthesis. The conversion process was monitored in real time by analyzing the intensity and full-width-at-half-maximum (FWHM) of the most prominent Raman peaks for classical TMDs' ($A_1'$) Raman mode and Janus ($A_1^1$) modes[62-64]. Fig.1a shows the time-dependent Raman spectrum of an exfoliated 1L-$WSe_2$ during its conversion to Janus $W_{Se}^{S}$. The overlaying characteristic $E' + A_1'$ modes[62,63] of 1L-$WSe_2$ at 254 cm$^{-1}$, shown in the red spectra in Fig. 1a, gradually disappear once the hydrogen plasma is ignited and the top chalcogen layer is etched. This Raman mode gradually broadens, indicating the reduction of crystallinity of 1L-$WSe_2$ and slowly blue-shifts to a higher frequency with increasing time, showing the strengthening of the partial W-Se bond. Furthermore, at ~ 90 seconds into the reaction, a new Raman mode at ~ 290 cm$^{-1}$ emerges and gradually redshifts and increases in intensity, shown by the overlaid spectra (bright orange to salmon pink) in Fig. 1a. between ~90 seconds and ~150 seconds. This mode is attributed to the $A_1^1$ Raman mode in Janus $WSe_{Se}^S$ [10,15,28,64]. Eventually, at time ~150 seconds the $E'+A_1'$ modes of 1L-WSe$_2$ are entirely diminished, and the $A_1^1$ Raman mode stabilizes at 284 cm$^{-1}$ frequency. The observed result is consistent with previous reports on 2D Janus conversion and the behaviour of the $A_1^1$ Raman mode is indicative of the relaxation of the partially etched and stiffened 1L-WSe$_2$ structure[28].

To gain insight into the microscopic structure of Janus monolayers during the Janus conversion process, HR-STEM was performed on a 1L-WSe$_2$ sample on a Quantifoil $^{TM}$ carbon grid at different stages of the Janus conversion, as shown in Fig. 1b to 1d. High Angle Annular Dark Field (HAADF) images on 1L-WSe$_2$, before the conversion, clearly shows the hexagonal lattice arrangement of the tungsten atom as bright spots and dimmer spots from two selenium atoms sitting on top of each other (Fig. 1b)[10,28]. The Rutherford scattering intensity from the tungsten and selenium atomic columns is plotted in the top panel of Fig. 1e (blue spectra) for the region shown by the blue line in Fig. 1b, where the arrow shows the direction from which the line profile in Fig. 1e is plotted.

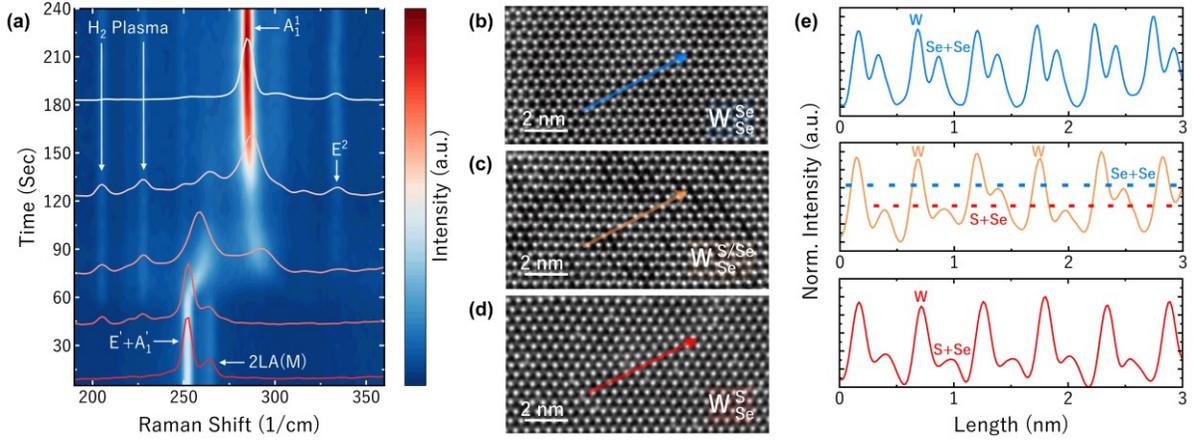

**Figure 1: Conversion of 1L-$WSe_2$ to Janus $W_{Se}^S$** - **(a)** Contour plot of time-dependent Raman spectra on 1L-$WSe_2$ to Janus $W_{Se}^S$: The red line overlay shows the $E'+A_1'$ Raman mode from 1L-$WSe_2$. The white line overlay shows the Janus $A_1^1$ mode in $W_{Se}^S$ **(b-d)** HR-STEM on 1L-$WSe_2$ at different times of conversions shown in Fig. 1a: **(b)** ) Shows HAADF micrograph of 1L-$WSe_2$, the structure is representative of the Raman spectra obtained before the ignition of plasma in Fig. 1a. **(c)** HAADF micrograph of partially converted 1L-$WSe_2$ to Janus $W_{Se}^S$ showing patches of Janus embedded into the parent TMDs, structure representative of the Raman spectra obtained between 60 to 150 seconds in Fig. 1a **(d)** HAADF micrograph of fully converted Janus $W_{Se}^S$, representative of the Raman spectra obtained at 240 seconds of reaction. **(e)** Corresponding line profile of the HAADF micrographs in Fig 1b, 1c and 1d: (Blue) Line profile of 1L-$WSe_2$ shown in Fig. 1b. (Orange) line profile of partially converted Janus + TMDs structure shown in Fig. 1c. (Red) line profile of fully converted Janus $W_{Se}^S$ in Fig. 1d.

The Z contrast in Fig.1b, from a single tungsten atom (Z=74), is comparable to two selenium atoms (Z=2*34) and is consistent with previous reports[15,17,18,28]. To capture the intermediate state of the Janus conversion (between 60 seconds to 150 seconds), the sample is converted for that duration on a Quantifoil grid supported on a Silicon substrate to prevent the bottom chalcogen layer from being etched. Fig. 1c shows the HAADF micrograph of the partially converted 1L-WSe$_2$ structure, where the conversion process is stopped after ~60 to 150 seconds. The orange curve in the middle panel of Fig. 1e shows the Z-contrast intensity extracted along the orange line in the HAADF micrograph in Fig.~1c. As the replacing chalcogen sulphur (Z=16) is lighter than the parent chalcogen selenium (Z=34), the overall intensity of the Janus chalcogen column decreases when compared with the intensity of the chalcogen column in the parent TMDs[65,66]. As a result, regions with dimmer spots can be observed in the HAADF micrograph. The sample is further subjected to the Janus conversion process until reaching the point of complete Janus conversion after ~240 seconds Fig. 1d (red line) shows the characteristic HAADF micrograph of the Janus – $W_{Se}^{S}$ structure, where the intensity of the chalcogen columns is significantly and homogeneously reduced compared to the parent WSe$_2$ [28].

The Z-contrast scattering intensity of the transition metal and chalcogen atomic columns for the $W_{Se}^{S}$ is plotted in the bottom panel of Fig. 1e. The selenium atoms (Z=34) have undergone a complete replacement by the lighter sulphur atoms (Z=16), resulting in the diminishing intensity when compared with the parent TMDs as shown in Fig. 1b. Although Fig. 1d shows a high degree of crystallinity, it is not representative the true crystal quality of the resulting Janus TMDs. This is because, after the conversion process, microscopic regions with a relative abundance of selenium impurities remain in the Janus material [10,28]. By comparing the Raman intensity of out-of-plane mode $A'_1$ of the parent TMDs before and after Janus conversion, one can gauge the conversion efficiency of the Janus conversion process [15,28]. All the samples studied in this letter showed above 98.3% conversion efficiency. The total reaction time required to achieve this efficiency was approximately 7 minutes and 30 seconds, including purge and vacuum cycles. This suggests that the etching step in the reaction is swift, as seen from the contour plot in Fig 1a; however, the chalcogen replacement dominates the total conversion time and is the rate-determining step, consistent with the recent study[67]. Additionally, HAADF simulations of monolayer WSe$_2$ and Janus $W_{Se}^{S}$ (See SI Fig. S1) are compared with the experimental data in Fig. 1b and 1d, the simulations match with the experimental datasets, and the overall findings from the HAADF micrographs and our Raman spectra indicate the presence of residual chalcogen atoms from the parent TMDs, vacancies, and point defects even within the most crystalline Janus materials [10,28].

**Identification of point defects in Janus $W_{Se}^{S}$.**

HR-STEM with a large Field of View (FOV) on $W_{Se}^{S}$ shows the presence of various types of point defects selected from highly defective regions (Fig. 2a). While other sections of the 2D Janus $W_{Se}^{S}$ do not contain a high density of defects, the selected region in Fig.2a closely captures all the four commonly observed energetically stable defects (total 7) across other fabricated samples. In a typical sample, four main types of point defects have been identified and indicated with circles: single chalcogen vacancies $V_S$ and $V_{Se}$ (red), double chalcogen vacancy $V_S-V_{Se}$ (pink); interstitial defect M$_i$ (light blue) and metal impurity M$_w$ (dark blue) consistent with defects observed for classical TMDs [20,35,40,68-70] except $V_S-V_{Se}$ double vacancies.

High magnification HAADF STEM image of single chalcogen vacancy is shown in Fig. 2b. The red arrow indicates the direction of the line profile, which is plotted in Fig. 2f and shown schematically in the right part of that figure. The HAADF STEM intensity provides distinguishable evidence of the absence of the two different chalcogen atoms due to the difference in their atomic number (Z). The top (bottom) panel shows the HAADF STEM intensity of a region with a sulphur (selenium) vacancy. Figure 2c shows the HAADF STEM image of a region with a double chalcogen vacancy (marked by a pink arrow). The line intensity profile is plotted in Fig. 2g and schematically presented in Fig 2g (left). In addition to the chalcogen vacancies, Fig. 2d also shows the presence of interstitial atoms (shown by the bright spot next to the centre of the light blue arrow); the line profile in Fig. 2h depicts the presence of an additional atom on the chalcogen site. Lastly, a bright atom is observed in Fig. 2e and shown within a dark blue arrow. This point defect is attributed to a metal impurity atom in the tungsten site, as depicted by the higher intensity in the line profile in Fig. 2i and schematically shown to the right.

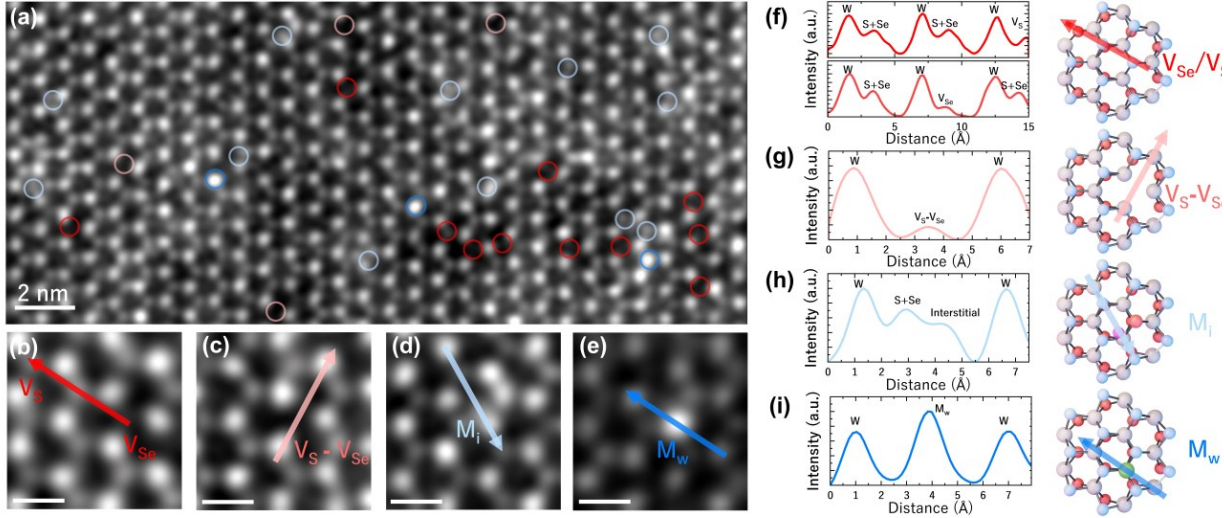

**Figure 2: Identification of point defects in 1L-Janus $W^S_{Se}$** - **(a)** HAADF image of 1L- Janus $W^S_{Se}$, circles show typical defects: single chalcogen vacancy, $V_S$ and $V_{Se}$ (red circle); double chalcogen vacancy, $V_S-V_{Se}$ (pink circle); interstitial defect, $M_i$ (light blue circle) and metal impurity, $M_w$ (dark blue circle) *(scalebar-5Å)*. **(b)** HAADF image of $V_S$ and $V_{Se}$ taken from a region in Fig. 2a. The arrowhead (red) indicates the direction from which the line profile (red) in **(f)** is constructed. The right part of (f) shows the schematic representation of the HAADF image in Fig. 2b. Similarly, **(c)** shows the HAADF image, and **(g)** shows the line profile (pink arrow) together with the schematic representation (right side) for the $V_S-V_{Se}$ **(d)** shows the HAADF image, and **(h)** shows Mi's line profile (light blue). Lastly, **(e)** shows the HAADF image, and **(i)** shows the corresponding line profile (dark blue) with schematic representation for a $M_w$

Here, single chalcogen $V_{Se}$ can be associated with defects in parent CVD or exfoliated WSe$_2$ or the formation of double selenium vacancy during plasma etching followed by sulphur decoration during the conversion process, yielding $V_{Se}$[28]. Similarly, the lighter chalcogen vacancy (sulphur) can be attributed to an incomplete or partial replacement process[17]. We note that the concentration of the double chalcogen vacancy is significantly smaller than that of the single chalcogen vacancy. This is likely related to the initial crystal quality and the defect activation energy of the double chalcogen vacancy during the Janus process [28]. Additionally, HAADF STEM simulations on the chalcogen vacancies are shown in SI Fig. S2 and are consistent with the experimental data in Fig. 2b and 2c. In contrast, observed $M_i$ and $M_w$ metal impurities are likely carried from the transition metal oxide or transition metal precursors used during CVD and CVT synthesis, respectively [33,71], and these defects are repeatedly observed in Janus monolayers from exfoliated and CVD-grown samples. Based on our observations, a higher number of single chalcogen vacancies, in particular, $V_S$ is observed and is attributed to the Janus conversion for even the most crystalline sample presently reported in literature [10], suggesting that even the best-in-class Janus sample has a large abundance of defects[10,28].

**One-dimensional (1D) defects in CVD Grown Janus Samples:**

Unlike Janus TMDs realized from exfoliated classical TMDs, CVD samples exhibit additional one-dimensional defects, suggesting an overall lower crystal quality of the parent TMD when compared to exfoliated monolayers [68,72,73] as depicted in Fig.3a[34,74-76]. As an example, the HAADF STEM micrograph of Janus $Mo^S_{Se}$ monolayers in Fig. 3a clearly show 1D line defects and grain boundaries (GBs) in addition to the previously discussed point defects. These 1D defects have been attributed to the polycrystalline nature of the parent CVD thin films[28,34,68] and elevated synthesis temperature in an uncontrolled reaction environment compared to bulk crystal growth techniques such as CVT and Flux methods [68,72,73,76].

Fig. 3b presents a high magnification region to show an example of a grain boundary in Janus $Mo^S_{Se}$ sample. The red box indicates the line profile, plotted in Fig. 3d, with a line defect schematically shown on the right side. This defect consists of chalcogen sites that have not undergone any etching and retains the parent chalcogen atom. This can be attributed to significantly different bonding characteristics around GBs compared to the single crystal parent TMDs requiring different process parameters (RF power) for their Janus conversion. This observation is consistent with established literature and suggests that precise control over the microstructural kinetics in the GB regions during the conversion process could yield higher conversion efficiency for polycrystalline samples [68,73]. Notably, a higher density of defects, commonly found in Janus TMDs converted from CVD samples, might be rearranging themselves to form these 1D defects outlined in Fig.3.

In stark contrast, the behaviour of 2D line defects shown in Fig. 3c is opposite to what is observed for the GBs in Fig 2b. The HAADF STEM and the subsequent line profile (blue box) show that these regions' chalcogen sites are terminated with the replacement chalcogen atom (sulphur). The chalcogen atom sites in the line defect have converted completely, suggesting a lower bond energy in this region compared to the parent matrix and GBs [28,34,68]. This indicates that optimizing the Janus conversion process for a higher yield and crystal quality in the case of any CVD sample will likely compromise the power tuning for conversion between these two regions. Ultimately, reducing the overall number of higher-order defects and limiting them to one type will likely contribute the most towards obtaining higher crystal quality in Janus TMDs

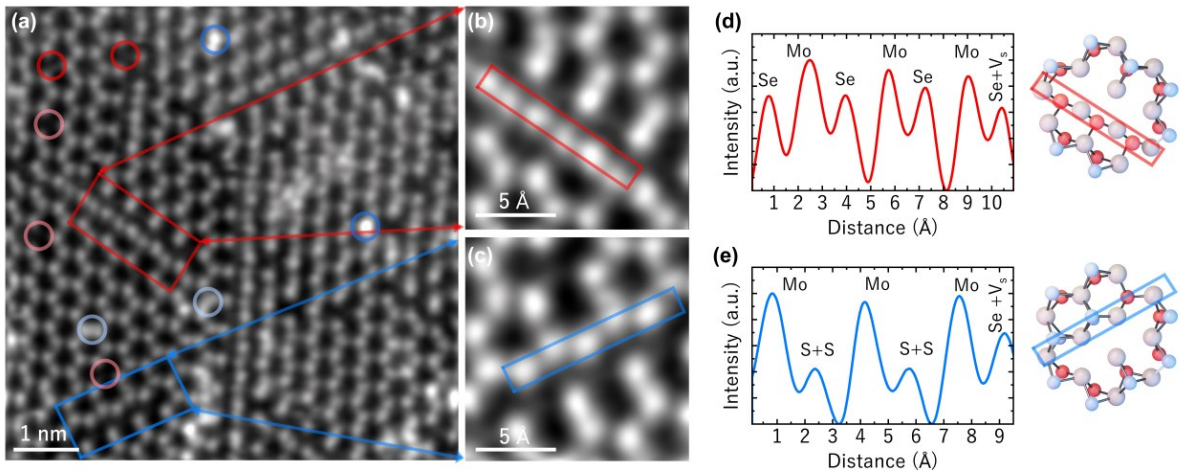

**Figure 3: Higher order defects in Janus $Mo^S_{Se}$ - (a)** HAADF image of Janus $Mo^S_{Se}$ synthesized from CVD-grown monolayers that show higher-order line and grain boundary defects. **(b)** HAADF image of GBs (red box) that was observed in Janus $Mo^S_{Se}$, their line profile analysis **(d)** shows that the chalcogen sites in the line defects are terminated with unetched parent chalcogen atoms, and the right part of (d) shows the schematic representation of the structure observed in Fig. 3b. Similarly, **(c)** HAADF image of a line defect (blue box) observed in Janus $Mo^S_{Se}$. Unlike the GBs, the chalcogen sites are entirely replaced by sulphur atoms **(e)** shows the line profile of the line defect in Fig. 3c along with the schematic representation of the crystal structure (right part)

**Theoretical Insights into Defect-Property Relations:**

First-principles computations with hybrid functional PBE0 were performed to study the thermodynamic stabilities of the relevant defects in Janus $W_{Se}^{S}$, including simple vacancy $V_S$, $V_{Se}$, $V_W$, and complex defects $V_S$–$V_{Se}$ and $V_S$–$V_{Se}$–$V_W$, as summarised in Fig 4. We have used a recently proposed scheme ensuring a proper representation of the band edges while enforcing 'Koopman's condition for the defects levels (see methods)[77]. The simulated defect configurations can be found in SI Fig. S3. Fig. 4a shows the defect formation energy versus Fermi level for all the defects in different charge states. It was found that among the simple defects, the chalcogen vacancies have the lowest formation energy, with $V_S$ slightly more stable than $V_{Se}$. All the defects that involve $V_W$ show a high formation energies (see SI Fig. S4), as shown in previous studies of metal vacancies in TMDs [78-80]. We now turn our attention to the interaction between chalcogen vacancies. Our results show that $V_S$–$V_{Se}$ vacancy complex is slightly more stable compared to non-interacting pairs of simple chalcogen vacancies with a binding energy of around 50 meV, as shown in the inset of Fig. 4a. We further note that the neutral charge state of $V_S$, $V_{Se}$, and $V_S$–$V_{Se}$ are the most relevant ones as they span the majority energy range in the band gap. With the most relevant defects identified, we plotted the electronic structure of $V_S^0$, $V_{Se}^0$ and $V_S$–$V_{Se}^0$, as shown in Fig. 4b.

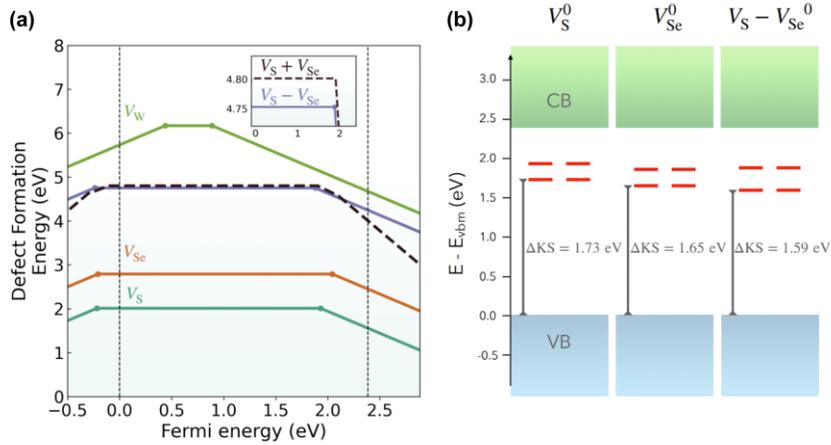

**Figure 4: First-principles calculations of defect levels in 1L Janus $W_{Se}^{S}$** – **(a)** defect formation energy diagram for intrinsic single and complex defects with charge states -1, 0, and +1 (solid lines). Dashed lines indicate the total energy of two single defects that compete with the formation of complex defects. $V_S$, $V_{Se}$ and $V_S - V_{Se}$ are favourable in a neutral charge state within the bandgap. The inset shows a zoomed-in plot illustrating that forming the complex defect $V_S - V_{Se}$ is slightly more favourable with the presence of $V_S$ and $V_{Se}$. We use the elemental chemical potential for each atom. **(b)** Single particle Kohn-Sham energy levels for the three most favourable defects at neutral charge state.

The Kohn-Sham level diagrams show that all three defects have four degenerate defect levels originating from the W dangling bond. We found that the chalcogen defects in Janus share a similar electronic structure compared with WS$_2$ and WSe$_2$[77,81,82]. Further incorporation of spin-orbit coupling (SOC) splits the defect levels by 210, 214, and 293 meV for $V_S^0$, $V_{Se}^0$ and $V_S$–$V_{Se}^0$, respectively (see SI Fig. S5 for SOC effect). We note that an optical excitation of these defects will involve the excitation of an electron in the valence band to the unoccupied defect state, forming a defect-bound exciton. We approximate the excitation energy based on the Kohn-Sham energy difference between the lowest unoccupied defect level and band edges ($\Delta KS$), which are 1.73 eV, 1.65 eV and 1.59 eV for $V_S^0$, $V_{Se}^0$ and $V_S$–$V_{Se}^0$, respectively. This computed trend will be used to understand the experimental PL spectrum discussed in the following section.

**Correlation of Bound Exciton Emission to Defect States:**

Our previous studies showed that significantly brighter emission in CVD Janus samples on SiO$_2$/Si substrates are related to bound exciton emission instead of previously identified neutral excitons[10,15,28]. To correlate defect states to emission characteristics, fully encapsulated h-BN/ W$_{Se}^{S}$/ h-BN heterostructures were fabricated to ensure no effective charge transfer takes place between SiO$_2$ and Janus monolayers and to reduce the exciton linewidth[60,61]. Fig. 5a shows the optical image of the fabricated Janus/h-BN heterostructure consisting of Janus W$_{Se}^{S}$ sandwiched between two separate few-layer h-BN layers on a silicon substrate with 300 nm thick thermal oxide. Monolayer WSe$_2$ was exfoliated and deterministically transferred onto the h-BN/SiO$_2$/Si, followed by Janus conversion to yield W$_{Se}^{S}$ and the Janus monolayer was fully encapsulated by another few-layer h-BN, as depicted in Fig. 5b. The room-temperature PL spectrum from fully encapsulated W$_{Se}^{S}$ (Fig. 5c) reveals a broad emission at 1.55 eV, a shoulder at 1.69 eV, and a distinct peak at 1.84 eV.

We note that the PL emission characteristics show significant differences between encapsulated and not encapsulated W$_{Se}^{S}$ as shown in Fig. 5c inset. Without encapsulation, W$_{Se}^{S}$ exhibits a single emission peak at 1.84 eV, attributed to the neutral exciton emission[10,28]. The prominence of this peak in the non-encapsulated device compared to the encapsulated device can be understood considering the charge transfer between the SiO$_2$ interface and the Janus TMDs. In the non-encapsulated device at room temperature, the defect levels are occupied due to effective doping from the substrate [10,63,83], which suppresses bound exciton emission. This is schematically shown in Fig. 5d. Upon encapsulation with h-BN, no effective charge transfer can occur, enabling the defect states to participate in exciton recombination pathways and give rise to the bound exciton emission.

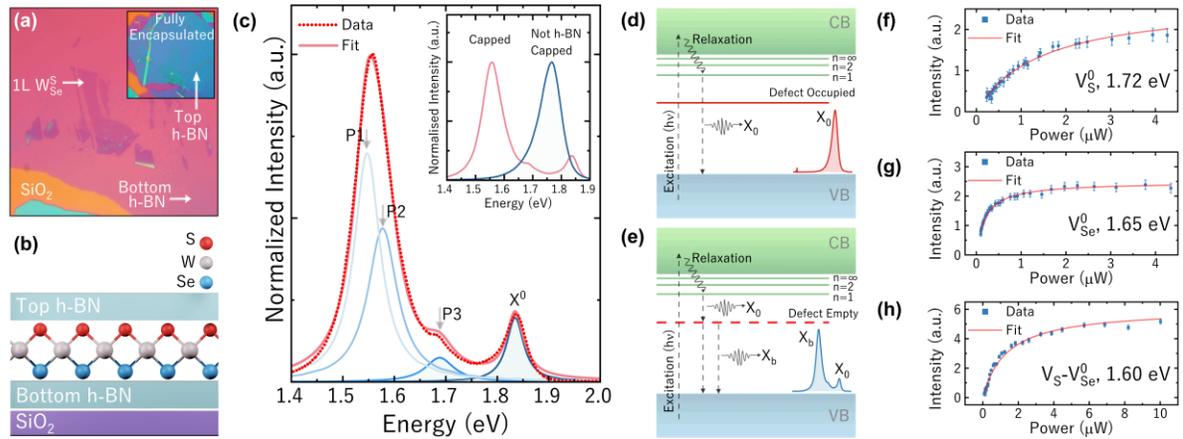

**Figure 5: Defect characterization of 1L Janus W$_{Se}^{S}$ encapsulated by h-BN and probed using cryogenic optical spectroscopy - (a)** Optical image of h-BN encapsulated 1L Janus W$_{Se}^{S}$. **(b)** Shows the schematic representation of the encapsulated Janus sample. **(c)** PL spectra of the fully encapsulated Janus monolayer show a substantial defect-related emission (solid red spectra), along with the characteristic neutral exciton emission X$^0$ at 1.84 eV at room temperature and additional peaks indicated by grey arrows. The cumulative peak fit (dotted line) is shown with its deconvolution, where three additional peaks labelled P1, P2 and P3 are identified (blue lines show their contribution). **(inset)** depicts a comparison between a non-encapsulated sample on SiO$_2$ and a fully hBN-encapsulated sample. **(d)** and **(e)** show the filling of defect levels associated with the charged transfer from the SiO$_2$ substrate to the non-encapsulated sample and the presence of unoccupied defect states when the sample is between hBN layers. **(f, g and h)** shows power-dependent PL spectra at 4K of the three peaks, P1, P2 and P3 identified in spectra (c).

Moreover, we note that the intensity of the defect-related emission bands is significantly more prominent compared to unbound excitons in the hBN-encapsulated Janus $W_{Se}^{S}$ device, suggesting that bound-exciton emission is the dominant path for exciton recombination in this device. The difference in the emission energy and intensity of the bound and neutral excitons in encapsulated and non-encapsulated samples is likely related to the combination of various phenomena, namely, physisorption of Oxygen ($O_2$) and moisture ($H_2O$), different effective dielectric screening from h-BN, and strain within the Janus monolayer[84-86]. In the case of the molecular physisorption process, which can be considered only for non-encapsulated samples, the monolayer Janus TMDs retain the p-type behaviour of the parent $WSe_2$, where the adsorption from oxygen and moisture depletes the sample from electrons, effectively increasing the trion emission and reducing the neutral exciton emission [84,86]. The significant shift from the neutral exciton at 1.84 eV (red spectra in the inset of Fig. 5c) to the highest located peak at 1.78 eV (blue spectra) is partially associated with this effect since it differs from results reported by Feuer et al. [10], the mentioned difference is likely related to various dielectric environments and/or strains. Upon encapsulation, the sample is protected from molecular physisorption and as a result, the contribution from different excitonic complexes, including bound exciton defect states, will be unaffected and more visible[85,86].

The broad nature of the defect-related emission shown in Fig. 5c is likely to be a convolution of multiple peaks attributed to various point defects shown in Fig. 2. The deconvoluted PL spectra fitted by Lorentzian functions show the presence of at least three distinct peaks at room temperature indicated by the labels P1, P2, and P3 (blue-shaded curves). The results of the fitting procedure regarding the energy of the P1, P2, and P3 emission peaks are in excellent agreement with our DFT calculations (Fig. 4) and are attributed to $V_S$, $V_{Se}$, and $V_S-V_{Se}$ at 1.68, 1.57, and 1.54 eV respectively at ambient temperatures and pressures. At cryogenic temperature (4 K), emission from bound exciton species dominates the optical emission from this device, and spectrally narrow emission peaks can be individually identified (See SI Fig. S6). Low-temperature power-dependence PL of these defect states revealed a saturation behaviour consistent with the nature of bound exciton emission in the TMDs [28,63]. Fig. 5f, 5g, and 5h show, respectively, the power-dependence of PL signatures at 1.72, 1.65 and 1.60 measured at 4K in the device shown in Fig. 5a. The power-dependence data are fit to a saturation curve assuming two-level system, yielding fit values for the saturation powers below 1 uW for all three spectral lines. The PL saturation at low optical power and saturation behavior consistent with a two-level system indicates the presence of isolated defects within the detection volume.

**Conclusions.** Overall, electron microscopy and optical spectroscopy studies, together with DFT simulations, introduced and structurally characterized the most common type of 0D point and 1D line defects in 2D Janus TMDs monolayers and correlated them to their optical properties. Cryogenic excitonic studies show that most defect states ($V_S$, $V_{Se}$, and $V_S-V_{Se}$) are optically active and manifest themselves in various bound excitonic states. Comprehensive density functional theory studies offered thermodynamic insights into the structural properties of these imperfections and correlated their energetic states and photoexcited carrier relaxation processes to the observed emission characteristics. Overall, studies point out that even for the best-in-class Janus materials reported to date, a significant number of point defects exist, and more creative growth methods, such as atomic layer deposition or solution-based processing, are needed to further improve the quality of 2D Janus layers with fine atomic level control. Overall findings introduced the genome of defects library of 2D Janus TMDs and established the first correlation between structural imperfections and material properties as a guideline for future studies, especially towards their synthesis using and the development of electronic devices capitalizing on exotic/superior properties of these emergent semiconductors.

**Methods:**

***CVD Synthesis:*** Monolayer Transition Metal Dichalcogenides (TMDs) were synthesized directly on SiO$_2$/Si substrates using established parameters from the literature[15,17,28]. Initially, the substrates were meticulously cleaned using a piranha etchant, followed by double sonication in deionized water to eliminate any residual etchant, followed by further sonication in ethanol and isopropyl alcohol for five minutes and subjected to a 75 W oxygen plasma treatment for fifteen minutes. The synthesis was done in an Atmospheric Pressure Chemical Vapor Deposition (AP-CVD) system. This system comprises a one-inch inner diameter quartz tube within a single-zone tube furnace (Lindberg Blue/M). The tube was cleaned with soap and deionized water before use. Less than 50 mg of Tungsten (VI) Oxide (Sigma-Aldrich, ACS reagent, ≥99.5%)) was distributed in a 7 ml Coors Tek combustion boat for TMDs synthesis. A 50 μM Perylene-3,4,9,10-tetracarboxylic acid tetrapotassium salt (PTAS) solution was applied to one substrate to enhance nucleation. Selenium powder, used as the chalcogen source, was placed upstream relative to the gas flow, with a reaction zone temperature of approximately 300°C. The furnace was ramped up to 850 °C in sixteen minutes, maintaining an argon flow rate of 80 SCCM and introducing 4 SCCM of hydrogen at the peak temperature for five minutes. Post-synthesis, the furnace cooled naturally, with the hydrogen supply cut off, and argon flow increased to 200 SCCM. The reaction of MoSe$_2$ monolayers was carried similarly with Molybdenum (VI) Oxide (Sigma-Aldrich, ACS reagent, ≥99.5%) precursor, reaction temperature of 760 °C and a flow rate of 46 SCCM UHP Argon.

***CVT Synthesis and exfoliations*** Bulk WSe2 crystals, synthesized via chemical vapour transport 15,33, briefly, Tungsten foil (Alfa Aesar, 99.999% metal basis ) and selenium powder (Thermo Scientific, 200 mesh, 99.999%) were added to a quartz ampoule in stochiometric ratio by mole basis and heated to 1000 °C gradually in 72 hours to prepare a polycrystalline charge. Before growth, the transition metals were treated with dilute nitric acid to remove native oxides. The polycrystalline powder is pulverized in a ball mill for 24 hours and analyzed via Wavelength Dispersive Spectroscopy (WDS) to quantify the stoichiometry. This powdered WSe2, mixed with 90 mg Tellurium Tetrachloride(TeCl4) as a transport agent, is added to a quartz ampoule that is cleaned with hydrofluoric acid (HF) to remove any organic residues. The ampoule is loaded in a three-zone furnace, with the precursors set at a hot zone at 1100°C and the growth zone at 1000°C on the other side of the ampoule. Single crystal WSe2 was obtained after 100 hours of growth and was exfoliated onto silicon substrates with a 300 nm thermal oxide layer. The exfoliation process involved multiple steps, initially utilizing Nitto Blue tape to achieve the desired sample thickness. Subsequently, this thinned sample was transferred to a polydimethylsiloxane (PDMS) layer to minimize tape residue. This PDMS stamp was carefully pressed against an ultra-clean silicon substrate, which had been previously treated with 75 W oxygen plasma. To further improve the contact between the substrate and the stamp, the PDMS silicon substrate was placed in a vacuum chamber for ten more minutes, eliminating air pockets. After removing the stamp, the sample was assessed by photoluminescence spectroscopy, explicitly looking for the absence of indirect emission to identify the regions of interest accurately.

***In-Situ Selective Epitaxy and Replacement:*** The synthesis of 2D Janus TMDs is performed in a custom-built glass reaction chamber consisting of two glass arms serving as the inlet and outlet manifold. Ultrapure hydrogen is introduced through the inlet, with the chamber's pressure regulated and monitored by a pressure gauge and capacitance manometer attached to the manifolds. CVD-grown and mechanically exfoliated TMD monolayers, used as reaction precursors, are placed at the centre of the synthesis chamber. Upstream to the gas flow, a secondary chalcogen source (sulphur) is positioned on a quartz stage. The chamber's outlet manifold is connected to a two-stage rotary vane pump (Edwards RV12), maintaining a pressure of 300 mTorr throughout the synthesis. Before synthesis, the chamber is evacuated of air and held under vacuum for fifteen minutes. Hydrogen at a flow rate of 20 SCCM is then introduced into the chamber and stabilized for ten minutes before commencing synthesis. Around the outlet manifold, a copper coil is connected to an RF generator (SEREN 101R) and an impedance-matching network (custom-designed). Hydrogen plasma is ignited by supplying 20 W of power to the coil. The synthesis process is conducted under a Raman Optical Spectrophotometer (Renishaw In-Via Confocal Raman Microscope) equipped with a 488 nm laser, focused using a long-distance objective (20x Infinity Corrected Mitutoyo Plan Apo SL). Raman spectroscopic data is collected to assess the optimal conversion of the TMDs into Janus monolayers (See Fig. 1a). Adjustments to power and flow rate are made throughout the conversion process based on the in-situ Raman data collected during the measurement.

***h-BN Janus Heterostructure Fabrication:*** The h-BN Janus heterostructures were prepared by first exfoliating a few layers of h-BN onto a silicon substrate with a 300 nm thermally grown oxide layer. Before exfoliation, the substrates were cleaned using piranha solution and treated with plasma to eliminate organic residues. Subsequently, the substrate underwent annealing at 150 °C in an ultra-high vacuum (UHV) environment to enhance the adhesion of the h-BN to the $SiO_2$ substrate. This step was followed by a chloroform treatment to remove any polydimethylsiloxane (PDMS) residues. Next, a monolayer of $WSe_2$ was deterministically transferred onto the substrate using a PDMS stamp, a transfer station, and a set of micromanipulators. The substrate with the $WSe_2$ layer was then annealed again at 150 °C and cleaned with chloroform to improve contact and remove any remaining PDMS residue. The $WSe_2$ monolayer on h-BN was converted into Janus $W^S_{Se}$ through the SEAR process, followed by encapsulation with an additional few layers of h-BN prepared on PDMS. The fully encapsulated structure underwent a final round of annealing and cleaning before being characterized by optical measurements.

***Cryogenic Photoluminescence Spectroscopy:*** All measurements at 4 K are carried out in a closed-cycle cryostat (the AttoDRY1000 from Atto cube Systems AG). Excitation and light collection are facilitated through a custom-built confocal microscope operating in reflection geometry. This microscope has an apochromatic objective lens with a numerical aperture (NA) of 0.81 (model LT-APO/NIR/0.81, also from Atto cube Systems AG). For the photoluminescence (PL) measurements, continuous-wave excitation is employed using a 2.33 eV laser (Ventus 532, from Laser Quantum Ltd.). The excitation power dependence is measured on the sample, and the optical intensity is calculated based on the optical spot size, determined by the 0.81 NA of the objective lens.

***Scanning Transmission Electron Microscopy and Sample Preparation:*** 2D Transition Metal Dichalcogenide (TMD) monolayers were exfoliated onto an SI substrate with 300 nm oxide layer and converted to Janus TMDs using SEAR, a solution of Poly(Bisphenol A Carbonate) (PC) in chloroform was spin-coated onto the substrate and dried, forming a uniform film. The TMDs/PC assembly was transferred then using adhesive thermal release tape (Nitto Denko REVALPHA, TRT). This assembly was carefully placed onto a Quantifoil holey carbon grid using micromanipulators. A glass slide was used to facilitate handling and heating it to 90 °C, separating the TMDs/PC structure from the TRT. The PC film was dissolved in heated chloroform vapour, and the transferred monolayer was finally inspected under a microscope and through Raman spectroscopy to confirm the absence of chloroform residue.

The HAADF STEM images were collected on a Nion Hermes Ultra STEM 100 microscope operated at 100 keV acceleration voltage, the probe convergence angle was set to 33 mrad, a beam current on the detector was about 25 pA, and an inner collection semi-angle of 80 mrad on the annular detector. Before the measurement, the sample was baked at 160 °C for 16 hours to reduce contamination during the experiment. The images were filtered and denoised by deconvolution of the point spread function (PSF) of the electron beam from the raw image by the Richardson-Lucy deconvolution algorithm.

*First-Principles Computations*: Defect computations were performed using Vienna Ab-initio Simulation Package (VASP)[87,88] and the projector-augmented wave (PAW) method[89] with the global hybrid functional PBE0[90,91]. A 144-atom orthorhombic monolayer supercell simulated each charged defect within a 24 Å vacuum layer. The plane-wave basis cutoff energy was set to 400 eV, and only the Γ point was used to sample the Brillouin zone. The defect structures were optimized at a fixed volume until the force on each ion was less than 0.01 eV/Å. The defect formation energy for each charged defect was computed using PyCDT. For a general defect $X$ at charge state $q$, its formation energy $E_{form}[X^q]$ is given by:

$$E_{form}[X^q] = E_{tot}[X^q] - E_{tot}^{bulk} - \sum n_i \mu_i + qE_F + E_{corr},$$

where $E_{form}[X^q]$ and $E_{tot}[X^q]$ are the total energies of the supercell containing the defect and the pristine one, respectively. The third term accounts for the energy needed to exchange atoms with the chemical reservoirs when creating the defect, where $n_i$ specifies the number of atoms with species $i$ being added ($n_i > 0$) or removed ($n_i < 0$) with corresponding chemical potential $\mu_i$. The fourth term accounts for the electron exchange with the host material given by the Fermi level $E_F$, which is commonly referenced as the valence band maximum (VBM). The last term contains corrections for the finite-size effects, and we employ the methods from Komsa et al.[92,93] as implemented in the SLABCC[94]. The chemical potential of each atom is set to that of its elemental form. The band-edge position for $WS_2$ was computed in a primitive monolayer cell with a 12x12x12 k-point mesh, and a 22% Fock exchange was used for an improved description of the bandgap. Defect levels (red solid lines) are computed using 7% PBE0 functionals to fulfill the generalized Koopman's condition as detailed in our previous work [77]. Spin-orbit coupling is considered throughout the electronic-structure calculations unless otherwise specified. The single-particle Kohn-Sham level correction includes the charge corrections and the vacuum level alignment.

**Acknowledgements**

ST acknowledges primary support from DOE-SC0020653 (materials synthesis), NSF ECCS 2052527, DMR 2111812, and CMMI 2129412. The use of facilities within the Eyring Materials Center at Arizona State University is partly supported by NNCI-ECCS-1542160. ST acknowledges partial support from Lawrence Semiconductor Labs. The US Department of Energy, Office of Science, Basic Energy Sciences in Quantum Information Science, under Award Number DE-SC0022289, supports the theoretical part of the work. This research used resources of the National Energy Research Scientific Computing Center; a DOE Office of Science User Facility supported by the Office of Science of the US Department of Energy under Contract No. DE-AC02-05CH11231 using NERSC award BES-ERCAP0020966. JK acknowledges support from the Polish National Agency for Academic Exchange within the Bekker program. RK acknowledges support from NCN Poland under grant OPUS 2018/29/B/ST7/02135. MYS acknowledges help and support from Evgeny M. Alexeev.

# References

1. Cheng, Y., Zhu, Z., Tahir, M. & Schwingenschlögl, U. Spin-orbit–induced spin splittings in polar transition metal dichalcogenide monolayers. *Europhysics Letters* **102**, 57001 (2013).
2. Mak, K. F., Lee, C., Hone, J., Shan, J. & Heinz, T. F. Atomically Thin ${\mathrm{MoS}}_{2}$: A New Direct-Gap Semiconductor. *Physical Review Letters* **105**, 136805, doi:10.1103/PhysRevLett.105.136805 (2010).
3. Zhang, J. *et al.* Janus monolayer transition-metal dichalcogenides. *ACS nano* **11**, 8192-8198 (2017).
4. Lu, A.-Y. *et al.* Janus monolayers of transition metal dichalcogenides. *Nature nanotechnology* **12**, 744-749 (2017).
5. Yao, Q.-F. *et al.* Manipulation of the large Rashba spin splitting in polar two-dimensional transition-metal dichalcogenides. *Physical Review B* **95**, 165401, doi:10.1103/PhysRevB.95.165401 (2017).
6. Hu, T. *et al.* Intrinsic and anisotropic Rashba spin splitting in Janus transition-metal dichalcogenide monolayers. *Physical Review B* **97**, 235404, doi:10.1103/PhysRevB.97.235404 (2018).
7. Chen, J., Wu, K., Ma, H., Hu, W. & Yang, J. Tunable Rashba spin splitting in Janus transition-metal dichalcogenide monolayers via charge doping. *RSC advances* **10**, 6388-6394 (2020).
8. Dong, L., Lou, J. & Shenoy, V. B. Large in-plane and vertical piezoelectricity in Janus transition metal dichalchogenides. *ACS nano* **11**, 8242-8248 (2017).
9. Cui, C., Xue, F., Hu, W.-J. & Li, L.-J. Two-dimensional materials with piezoelectric and ferroelectric functionalities. *npj 2D Materials and Applications* **2**, 18, doi:10.1038/s41699-018-0063-5 (2018).
10. Feuer, M. S. *et al.* Identification of Exciton Complexes in Charge-Tunable Janus WSeS Monolayers. *ACS nano* **17**, 7326-7334 (2023).
11. Liang, J. *et al.* Very large Dzyaloshinskii-Moriya interaction in two-dimensional Janus manganese dichalcogenides and its application to realize skyrmion states. *Physical Review B* **101**, 184401 (2020).
12. Yuan, J. *et al.* Intrinsic skyrmions in monolayer Janus magnets. *Physical Review B* **101**, 094420 (2020).
13. Han, Y.-t., Ji, W.-x., Wang, P.-J., Li, P. & Zhang, C.-W. Strain-tunable skyrmions in two-dimensional monolayer Janus magnets. *Nanoscale* **15**, 6830-6837 (2023).
14. Guan, Z., Ni, S. & Hu, S. Tunable electronic and optical properties of monolayer and multilayer Janus MoSSe as a photocatalyst for solar water splitting: a first-principles study. *The Journal of Physical Chemistry C* **122**, 6209-6216 (2018).
15. Sayyad, M. *et al.* Strain Anisotropy Driven Spontaneous Formation of Nanoscrolls from 2D Janus Layers. *Advanced Functional Materials* **33**, 2303526, doi:https://doi.org/10.1002/adfm.202303526 (2023).
16. Kim, S. W. *et al.* Understanding Solvent-Induced Delamination and Intense Water Adsorption in Janus Transition Metal Dichalcogenides for Enhanced Device Performance.

*Advanced Functional Materials* **n/a**, 2308709, doi:https://doi.org/10.1002/adfm.202308709 (2023).
17  Trivedi, D. B. *et al.* Room-temperature synthesis of 2D Janus crystals and their heterostructures. *Advanced materials* **32**, 2006320 (2020).
18  Lin, Y.-C. *et al.* Low energy implantation into transition-metal dichalcogenide monolayers to form Janus structures. *ACS nano* **14**, 3896-3906 (2020).
19  Guo, Y. *et al.* Designing artificial two-dimensional landscapes via atomic-layer substitution. *Proceedings of the National Academy of Sciences* **118**, e2106124118, doi:10.1073/pnas.2106124118 (2021).
20  Gan, Z. *et al.* Chemical Vapor Deposition of High-Optical-Quality Large-Area Monolayer Janus Transition Metal Dichalcogenides. *Advanced Materials* **34**, 2205226, doi:https://doi.org/10.1002/adma.202205226 (2022).
21  Li, H. *et al.* Anomalous behavior of 2D Janus excitonic layers under extreme pressures. *Advanced Materials* **32**, 2002401 (2020).
22  Yagmurcukardes, M. *et al.* Quantum properties and applications of 2D Janus crystals and their superlattices. *Applied Physics Reviews* **7** (2020).
23  Zheng, T. *et al.* Excitonic Dynamics in Janus MoSSe and WSSe Monolayers. *Nano Letters* **21**, 931-937, doi:10.1021/acs.nanolett.0c03412 (2021).
24  Petrić, M. M. *et al.* Nonlinear Dispersion Relation and Out-of-Plane Second Harmonic Generation in MoSSe and WSSe Janus Monolayers (Advanced Optical Materials 19/2023). *Advanced Optical Materials* **11**, 2370076, doi:https://doi.org/10.1002/adom.202370076 (2023).
25  Shi, J. *et al.* Giant room-temperature nonlinearities in a monolayer Janus topological semiconductor. *Nature Communications* **14**, 4953, doi:10.1038/s41467-023-40373-z (2023).
26  Zheng, T. *et al.* Janus monolayers for ultrafast and directional charge transfer in transition metal dichalcogenide heterostructures. *ACS nano* **16**, 4197-4205 (2022).
27  Strasser, A., Wang, H. & Qian, X. Nonlinear Optical and Photocurrent Responses in Janus MoSSe Monolayer and MoS2–MoSSe van der Waals Heterostructure. *Nano Letters* **22**, 4145-4152, doi:10.1021/acs.nanolett.2c00898 (2022).
28  Qin, Y. *et al.* Reaching the excitonic limit in 2D Janus monolayers by in situ deterministic growth. *Advanced Materials* **34**, 2106222 (2022).
29  Li, F., Wei, W., Zhao, P., Huang, B. & Dai, Y. Electronic and Optical Properties of Pristine and Vertical and Lateral Heterostructures of Janus MoSSe and WSSe. *The Journal of Physical Chemistry Letters* **8**, 5959-5965, doi:10.1021/acs.jpclett.7b02841 (2017).
30  Ma, Y., Kou, L., Huang, B., Dai, Y. & Heine, T. Two-dimensional ferroelastic topological insulators in single-layer Janus transition metal dichalcogenides M SSe (M= Mo, W). *Physical Review B* **98**, 085420 (2018).
31  Liu, J. *et al.* Linear photogalvanic effects in Janus monolayer MoSSe with intrinsic defects. *Optics & Laser Technology* **159**, 108946 (2023).
32  Červenka, J., Katsnelson, M. I. & Flipse, C. F. J. Room-temperature ferromagnetism in graphite driven by two-dimensional networks of point defects. *Nature Physics* **5**, 840-844, doi:10.1038/nphys1399 (2009).
33  Tongay, S. *et al.* Defects activated photoluminescence in two-dimensional semiconductors: interplay between bound, charged and free excitons. *Scientific Reports* **3**, 2657, doi:10.1038/srep02657 (2013).
34  Gibb, A. L. *et al.* Atomic Resolution Imaging of Grain Boundary Defects in Monolayer Chemical Vapor Deposition-Grown Hexagonal Boron Nitride. *Journal of the American Chemical Society* **135**, 6758-6761, doi:10.1021/ja400637n (2013).

35  Lin, Z. *et al.* 2D materials advances: from large scale synthesis and controlled heterostructures to improved characterization techniques, defects and applications. *2D Materials* **3**, 042001 (2016).
36  Wu, Z. & Ni, Z. Spectroscopic investigation of defects in two-dimensional materials. *Nanophotonics* **6**, 1219-1237 (2017).
37  Zheng, Y. J. *et al.* Point defects and localized excitons in 2D WSe2. *ACS nano* **13**, 6050-6059 (2019).
38  Turiansky, M. E., Alkauskas, A. & Van de Walle, C. G. Spinning up quantum defects in 2D materials. *Nature Materials* **19**, 487-489, doi:10.1038/s41563-020-0668-x (2020).
39  Tian, X. *et al.* Correlating the three-dimensional atomic defects and electronic properties of two-dimensional transition metal dichalcogenides. *Nature Materials* **19**, 867-873, doi:10.1038/s41563-020-0636-5 (2020).
40  Mitterreiter, E. *et al.* The role of chalcogen vacancies for atomic defect emission in MoS2. *Nature Communications* **12**, 3822, doi:10.1038/s41467-021-24102-y (2021).
41  Xu, J. *et al.* Atomic-level polarization in electric fields of defects for electrocatalysis. *Nature Communications* **14**, 7849, doi:10.1038/s41467-023-43689-y (2023).
42  Xiao, D., Liu, G.-B., Feng, W., Xu, X. & Yao, W. Coupled Spin and Valley Physics in Monolayers of ${\mathrm{MoS}}_{2}$ and Other Group-VI Dichalcogenides. *Physical Review Letters* **108**, 196802, doi:10.1103/PhysRevLett.108.196802 (2012).
43  Mak, K. F., He, K., Shan, J. & Heinz, T. F. Control of valley polarization in monolayer MoS2 by optical helicity. *Nature Nanotechnology* **7**, 494-498, doi:10.1038/nnano.2012.96 (2012).
44  Jones, A. M. *et al.* Optical generation of excitonic valley coherence in monolayer WSe2. *Nature Nanotechnology* **8**, 634-638, doi:10.1038/nnano.2013.151 (2013).
45  Schaibley, J. R. *et al.* Valleytronics in 2D materials. *Nature Reviews Materials* **1**, 1-15 (2016).
46  Mak, K. F. & Shan, J. Opportunities and challenges of interlayer exciton control and manipulation. *Nature Nanotechnology* **13**, 974-976, doi:10.1038/s41565-018-0301-1 (2018).
47  Zeng, H., Dai, J., Yao, W., Xiao, D. & Cui, X. Valley polarization in MoS2 monolayers by optical pumping. *Nature Nanotechnology* **7**, 490-493, doi:10.1038/nnano.2012.95 (2012).
48  Wang, Z., Shan, J. & Mak, K. F. Valley- and spin-polarized Landau levels in monolayer WSe2. *Nature Nanotechnology* **12**, 144-149, doi:10.1038/nnano.2016.213 (2017).
49  Lyons, T. P. *et al.* The valley Zeeman effect in inter- and intra-valley trions in monolayer WSe2. *Nature Communications* **10**, 2330, doi:10.1038/s41467-019-10228-7 (2019).
50  Aivazian, G. *et al.* Magnetic control of valley pseudospin in monolayer WSe2. *Nature Physics* **11**, 148-152, doi:10.1038/nphys3201 (2015).
51  Luo, C., Peng, X., Qu, J. & Zhong, J. Valley degree of freedom in ferromagnetic Janus monolayer H-VSSe and the asymmetry-based tuning of the valleytronic properties. *Physical Review B* **101**, 245416 (2020).
52  He, Y.-M. *et al.* Single quantum emitters in monolayer semiconductors. *Nature Nanotechnology* **10**, 497-502, doi:10.1038/nnano.2015.75 (2015).
53  Palacios-Berraquero, C. *et al.* Large-scale quantum-emitter arrays in atomically thin semiconductors. *Nature Communications* **8**, 15093, doi:10.1038/ncomms15093 (2017).
54  Parto, K., Azzam, S. I., Banerjee, K. & Moody, G. Defect and strain engineering of monolayer WSe2 enables site-controlled single-photon emission up to 150 K. *Nature Communications* **12**, 3585, doi:10.1038/s41467-021-23709-5 (2021).
55  Andrei, E. Y. *et al.* The marvels of moiré materials. *Nature Reviews Materials* **6**, 201-206, doi:10.1038/s41578-021-00284-1 (2021).

56  Shabani, S. *et al.* Deep moiré potentials in twisted transition metal dichalcogenide bilayers. *Nature Physics* **17**, 720-725, doi:10.1038/s41567-021-01174-7 (2021).
57  Li, H. *et al.* Imaging moiré flat bands in three-dimensional reconstructed WSe2/WS2 superlattices. *Nature Materials* **20**, 945-950, doi:10.1038/s41563-021-00923-6 (2021).
58  Huang, D., Choi, J., Shih, C.-K. & Li, X. Excitons in semiconductor moiré superlattices. *Nature Nanotechnology* **17**, 227-238, doi:10.1038/s41565-021-01068-y (2022).
59  Verhagen, T., Guerra, V. L. P., Haider, G., Kalbac, M. & Vejpravova, J. Towards the evaluation of defects in MoS2 using cryogenic photoluminescence spectroscopy. *Nanoscale* **12**, 3019-3028, doi:10.1039/C9NR07246B (2020).
60  Cadiz, F. *et al.* Excitonic linewidth approaching the homogeneous limit in MoS 2-based van der Waals heterostructures. *Physical Review X* **7**, 021026 (2017).
61  Man, M. K. L. *et al.* Protecting the properties of monolayer MoS2 on silicon based substrates with an atomically thin buffer. *Scientific Reports* **6**, 20890, doi:10.1038/srep20890 (2016).
62  Li, Z. *et al.* Revealing the biexciton and trion-exciton complexes in BN encapsulated WSe2. *Nature Communications* **9**, 3719, doi:10.1038/s41467-018-05863-5 (2018).
63  Barbone, M. *et al.* Charge-tuneable biexciton complexes in monolayer WSe2. *Nature Communications* **9**, 3721, doi:10.1038/s41467-018-05632-4 (2018).
64  Petrić, M. M. *et al.* Raman spectrum of Janus transition metal dichalcogenide monolayers WSSe and MoSSe. *Physical Review B* **103**, 035414 (2021).
65  Pennycook, S. J. & Boatner, L. A. Chemically sensitive structure-imaging with a scanning transmission electron microscope. *Nature* **336**, 565-567, doi:10.1038/336565a0 (1988).
66  Yamashita, S. *et al.* Atomic number dependence of Z contrast in scanning transmission electron microscopy. *Scientific Reports* **8**, 12325, doi:10.1038/s41598-018-30941-5 (2018).
67  Suzuki, H. *et al.* Intermediate State between MoSe2 and Janus MoSeS during Atomic Substitution Process. *Nano Letters* (2023).
68  van der Zande, A. M. *et al.* Grains and grain boundaries in highly crystalline monolayer molybdenum disulphide. *Nature Materials* **12**, 554-561, doi:10.1038/nmat3633 (2013).
69  Zhang, T., Wang, J., Wu, P., Lu, A.-Y. & Kong, J. Vapour-phase deposition of two-dimensional layered chalcogenides. *Nature Reviews Materials* **8**, 799-821, doi:10.1038/s41578-023-00609-2 (2023).
70  Bertoldo, F. *et al.* Intrinsic Defects in MoS2 Grown by Pulsed Laser Deposition: From Monolayers to Bilayers. *ACS Nano* **15**, 2858-2868, doi:10.1021/acsnano.0c08835 (2021).
71  Zhang, J. *et al.* Characterization of atomic defects on the photoluminescence in two-dimensional materials using transmission electron microscope. *InfoMat* **1**, 85-97 (2019).
72  Yu, Q. *et al.* Control and characterization of individual grains and grain boundaries in graphene grown by chemical vapour deposition. *Nature Materials* **10**, 443-449, doi:10.1038/nmat3010 (2011).
73  Najmaei, S. *et al.* Vapour phase growth and grain boundary structure of molybdenum disulphide atomic layers. *Nature Materials* **12**, 754-759, doi:10.1038/nmat3673 (2013).
74  Karvonen, L. *et al.* Rapid visualization of grain boundaries in monolayer MoS2 by multiphoton microscopy. *Nature Communications* **8**, 15714, doi:10.1038/ncomms15714 (2017).
75  He, Y. *et al.* Engineering grain boundaries at the 2D limit for the hydrogen evolution reaction. *Nature Communications* **11**, 57, doi:10.1038/s41467-019-13631-2 (2020).
76  Reifsnyder Hickey, D. *et al.* Illuminating Invisible Grain Boundaries in Coalesced Single-Orientation WS2 Monolayer Films. *Nano Letters* **21**, 6487-6495, doi:10.1021/acs.nanolett.1c01517 (2021).

77  Chen, W., Griffin, S. M., Rignanese, G.-M. & Hautier, G. Nonunique fraction of Fock exchange for defects in two-dimensional materials. *Physical Review B* **106**, L161107, doi:10.1103/PhysRevB.106.L161107 (2022).

78  Singh, A. & Singh, A. K. Atypical behavior of intrinsic defects and promising dopants in two-dimensional $\mathrm{WS}_{2}$. *Physical Review Materials* **5**, 084001, doi:10.1103/PhysRevMaterials.5.084001 (2021).

79  Zhou, W. *et al.* Intrinsic Structural Defects in Monolayer Molybdenum Disulfide. *Nano Letters* **13**, 2615-2622, doi:10.1021/nl4007479 (2013).

80  Hong, J. *et al.* Exploring atomic defects in molybdenum disulphide monolayers. *Nature Communications* **6**, 6293, doi:10.1038/ncomms7293 (2015).

81  Barja, S. *et al.* Identifying substitutional oxygen as a prolific point defect in monolayer transition metal dichalcogenides. *Nature Communications* **10**, 3382, doi:10.1038/s41467-019-11342-2 (2019).

82  Schuler, B. *et al.* Large Spin-Orbit Splitting of Deep In-Gap Defect States of Engineered Sulfur Vacancies in Monolayer $\mathrm{WS}_{2}$. *Physical Review Letters* **123**, 076801, doi:10.1103/PhysRevLett.123.076801 (2019).

83  Iberi, V. *et al.* Nanoforging Single Layer MoSe2 Through Defect Engineering with Focused Helium Ion Beams. *Scientific Reports* **6**, 30481, doi:10.1038/srep30481 (2016).

84  Tongay, S. *et al.* Broad-Range Modulation of Light Emission in Two-Dimensional Semiconductors by Molecular Physisorption Gating. *Nano Letters* **13**, 2831-2836, doi:10.1021/nl4011172 (2013).

85  Mak, K. F. *et al.* Tightly bound trions in monolayer MoS2. *Nature Materials* **12**, 207-211, doi:10.1038/nmat3505 (2013).

86  Raja, A. *et al.* Coulomb engineering of the bandgap and excitons in two-dimensional materials. *Nature Communications* **8**, 15251, doi:10.1038/ncomms15251 (2017).

87  Kresse, G. & Furthmüller, J. Efficiency of ab-initio total energy calculations for metals and semiconductors using a plane-wave basis set. *Computational Materials Science* **6**, 15-50, doi:https://doi.org/10.1016/0927-0256(96)00008-0 (1996).

88  Kresse, G. & Furthmüller, J. Efficient iterative schemes for ab initio total-energy calculations using a plane-wave basis set. *Physical Review B* **54**, 11169-11186, doi:10.1103/PhysRevB.54.11169 (1996).

89  Blöchl, P. E. Projector augmented-wave method. *Physical Review B* **50**, 17953-17979, doi:10.1103/PhysRevB.50.17953 (1994).

90  Adamo, C. & Barone, V. Toward reliable density functional methods without adjustable parameters: The PBE0 model. *The Journal of Chemical Physics* **110**, 6158-6170, doi:10.1063/1.478522 (1999).

91  Perdew, J. P., Ernzerhof, M. & Burke, K. Rationale for mixing exact exchange with density functional approximations. *The Journal of Chemical Physics* **105**, 9982-9985, doi:10.1063/1.472933 (1996).

92  Komsa, H.-P., Berseneva, N., Krasheninnikov, A. V. & Nieminen, R. M. Charged Point Defects in the Flatland: Accurate Formation Energy Calculations in Two-Dimensional Materials. *Physical Review X* **4**, 031044, doi:10.1103/PhysRevX.4.031044 (2014).

93  Komsa, H.-P. & Pasquarello, A. Finite-Size Supercell Correction for Charged Defects at Surfaces and Interfaces. *Physical Review Letters* **110**, 095505, doi:10.1103/PhysRevLett.110.095505 (2013).

94  Farzalipour Tabriz, M., Aradi, B., Frauenheim, T. & Deák, P. SLABCC: Total energy correction code for charged periodic slab models. *Computer Physics Communications* **240**, 101-105, doi:https://doi.org/10.1016/j.cpc.2019.02.018 (2019).

# The Defects Genome of Janus Transition Metal Dichalcogenides

Mohammed Sayyad [Γ,1], Jan Kopaczek [2], Carmem M. Gilardoni [3], Weiru Chen [4], Yihuang Xiong [4], Shize Yang [5], Kenji Watanabe [6], Takashi Taniguchi [7], Robert Kudrawiec [2], Geoffroy Hautier [4], Mete Atatüre[3] and Sefaattin Tongay [Γ, 1]

[1] Materials Science and Engineering, School for Engineering of Matter, Transport and Energy, Arizona State University, Tempe, Arizona, AZ 85287, United States of America.

[2] Department of Semiconductor Materials Engineering, Faculty of Fundamental Problems of Technology, Wroclaw University of Science and Technology, Wybrzeże Stanisława Wyspiańskiego 27, 50-370 Wroclaw, Poland.

[3] Cavendish Laboratory, University of Cambridge, J.J. Thomson Avenue, Cambridge, CB3 0HE, United Kingdom.

[4] Thayer School of Engineering, Dartmouth College, Hanover, New Hampshire, NH 03755, United States of America.

[5] Aberration Corrected Electron Microscopy Core, Yale University, New Haven, Connecticut, CT 06516, United States of America.

[6] Research Center for Functional Materials, National Institute for Materials Science, Tsukuba 305-0044, Japan.

[7] International Center for Materials Nanoarchitectonics, National Institute for Materials Science, Tsukuba305-0044, Japan.

[Γ] *Corresponding Authors: sefaattin.tongay@asu.edu, msayyad@asu.edu*

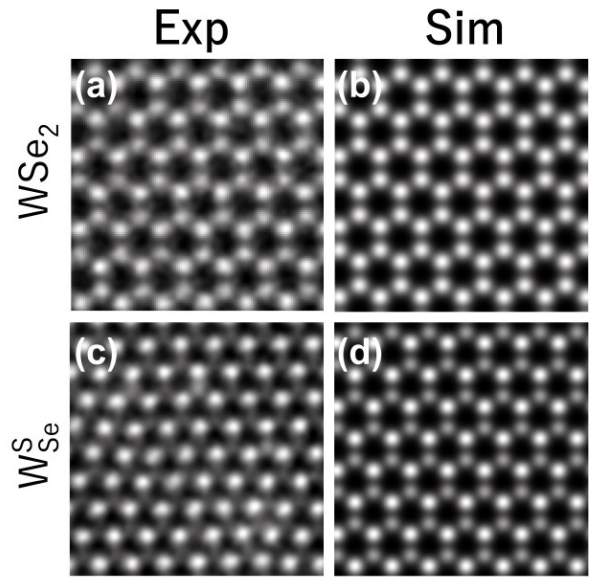

**Figure S1: Simulated HAADF in 1L WSe$_2$ and Janus $W^S_{Se}$ – (a)** Experimental HAADF micrograph and **(b)** simulated HAADF micrograph of 1L- WSe$_2$ similarly, **(c)** shows the Experimental HAADF micrograph of 1L-Janus $W^S_{Se}$ and **(d)** shows its simulated HAADF micrograph.

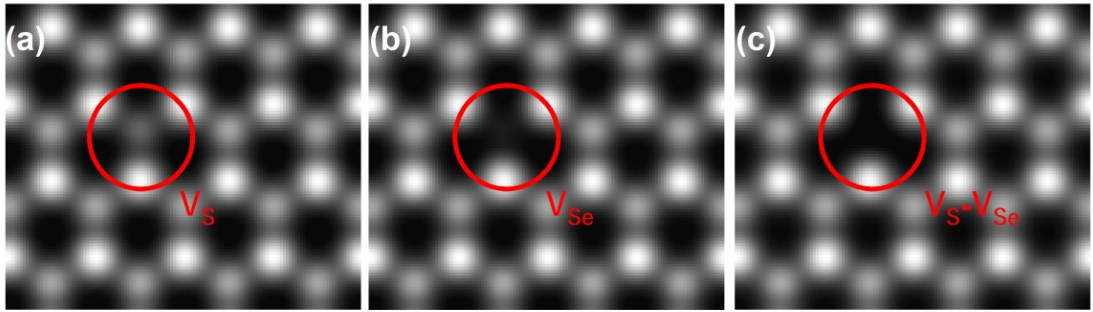

**Figure S2:** Simulated HAADF micrographs of defect configurations in 1L Janus $W^S_{Se}$– **(a)** single sulfur vacancy $V_S$, **(b)** single selenium vacancy $V_{Se}$, **(c)** sulfur-selenium vacancy $V_S$–$V_{Se}$,

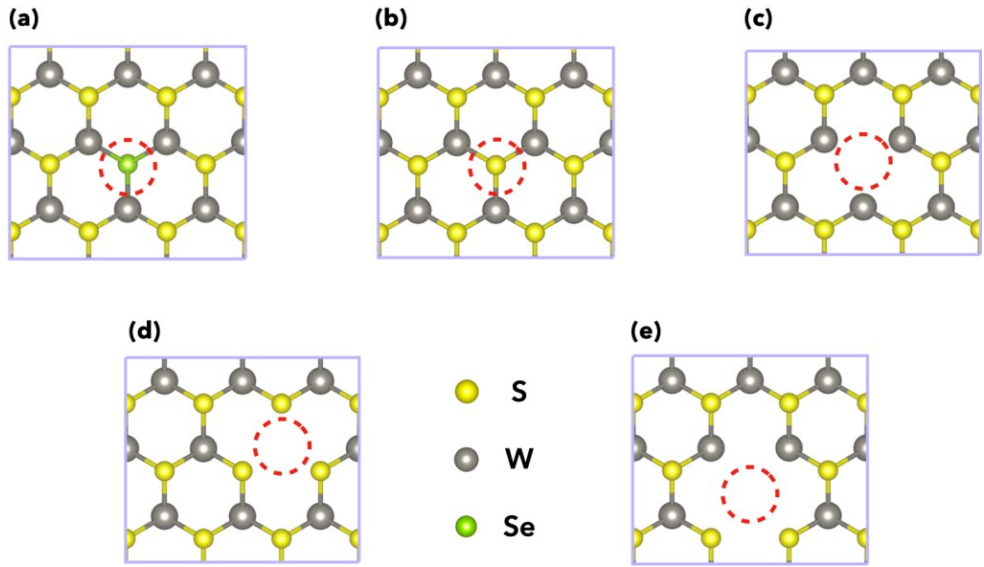

**Figure S3:** Simulated defect configurations in 1L Janus $W^S_{Se}$– **(a)** single sulfur vacancy $V_S$, **(b)** single selenium vacancy $V_{Se}$, **(c)** single tungsten vacancy $V_W$, **(d)** sulfur-selenium vacancy $V_S$–$V_{Se}$, and **(e)** sulfur-seleninum-tungsten vacancy $V_S$–$V_{Se}$–$V_W$.

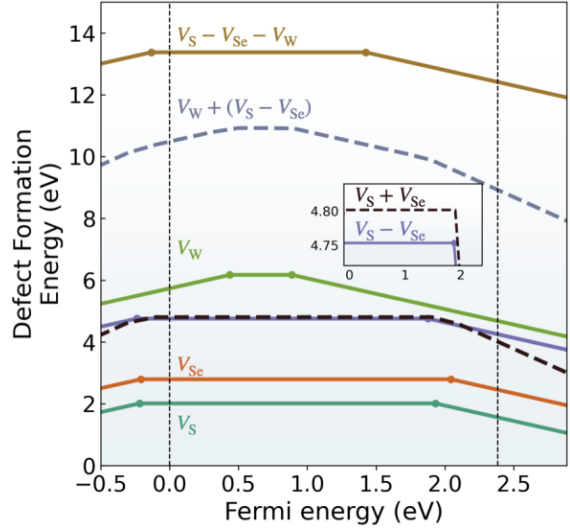

**Figure S4: Defect formation energy diagram in 1L Janus $W^S_{Se}$** – formation energy diagram for selected simple and complex defects at charge state –1, 0 and +1. Dashed lines indicate the competition between forming complex defects and simple defects.

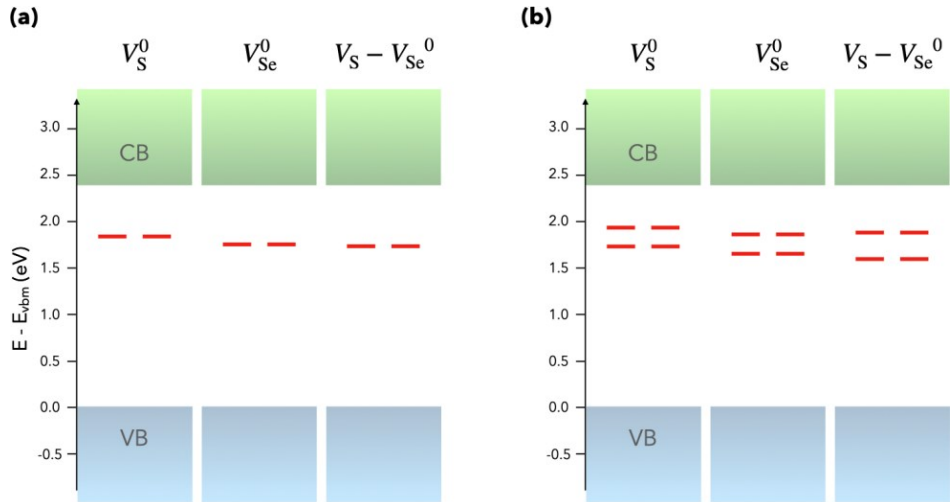

**Figure S5: Kohn-Sham energy level diagrams for neutral $V_S$, $V_{Se}$ and $V_S$-$V_{Se}$.** (a) Without SOC, the ground state is a singlet with 4 degenerate states within the bandgap. Here only one spin channel is shown for simplicity. (b) SOC splits the degenerate state by 210 meV, 214 meV and 293 meV for neutral $V_S$, $V_{Se}$ and $V_S$-$V_{Se}$, respectively.

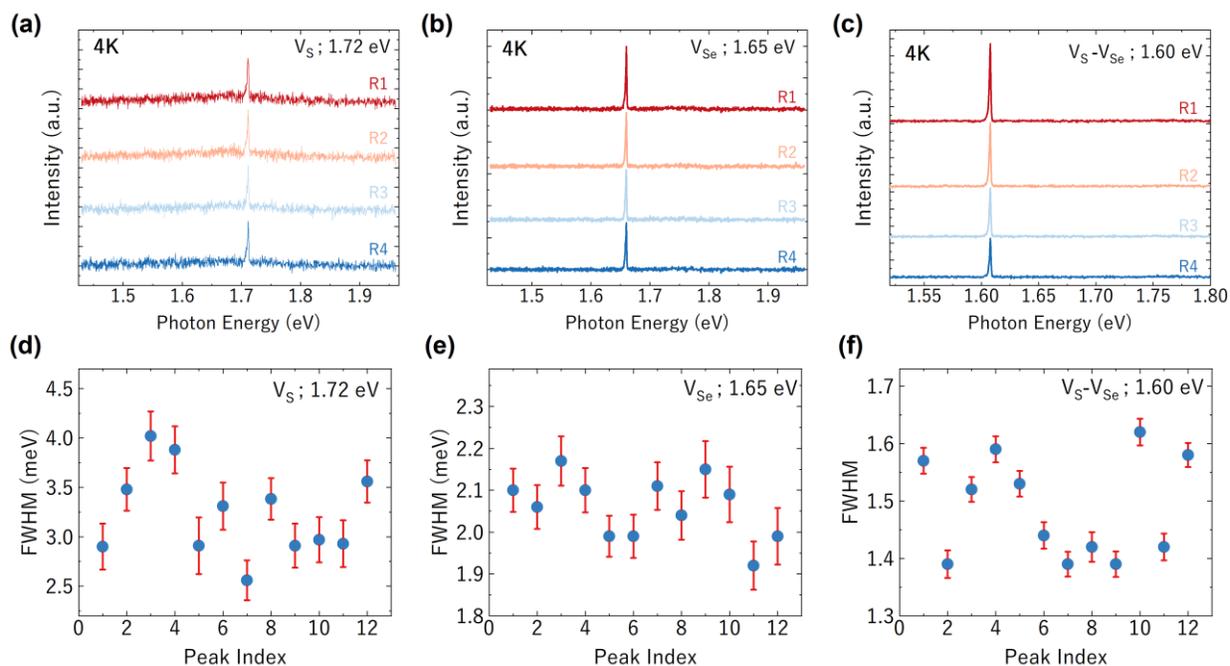

**Figure S6: Cryogenic PL Spectra in 1L Janus $W_{Se}^{S}$ device shown in Fig. 51** – **(a)** PL spectra obtained from four random spots at 1.72 eV attributed to single sulfur vacancies $V_S$, similarly **(b)** shows the PL emission from four random spots at 1.65 eV attributed to single selenium vacancy $V_{Se}$, **(c)** shows PL spectra obtained at 1.60 eV attributed to sulfur-selenium vacancy $V_S - V_{Se}$, from four random spots: All spectra were processed and analysed via a MATLAB script, with specified wavelength limits as a guide for peak search algorithm. The algorithm implements a Lorentzian profile fitting to identify the relevant peaks and extract critical parameters, ensuring precise spectral characterization **(d)** Shows the distribution of the FWHM for peak at 1.72 eV for twelve random PL spectra. Similarly **(e)** and **(f)** show the distribution of FWHM for the peaks at 1.65 eV and 1.60 eV respectively.